\documentclass[draft,nofootinbib]{revtex4}

\begin{document}

%
%

\title{DRAKE EQUATION FOR THE MULTIVERSE: FROM THE STRING LANDSCAPE TO COMPLEX LIFE}

\author{MARCELO GLEISER}

\affiliation{Department of Physics and Astronomy, Dartmouth College\\
Hanover, NH 03755, USA\\
mgleiser@dartmouth.edu}

\begin{abstract}
It is argued that selection criteria usually referred to as ``anthropic conditions'' for the existence of intelligent (typical) observers widely adopted in cosmology amount only to preconditions for primitive life. The existence of life does not imply in the existence of intelligent life. On the contrary, the transition from single-celled to complex, multi-cellular organisms is far from trivial, requiring stringent additional conditions on planetary platforms. An attempt is made to disentangle the necessary steps leading from a selection of universes out of a hypothetical multiverse to the existence of life and of complex life. It is suggested that what is currently called the ``anthropic principle'' should instead be named the ``prebiotic principle.'' 
\end{abstract}

\maketitle


\section{Introduction}	

The hypothetical existence of a multiverse, be it inspired by the string landscape \cite{Susskind} or by cosmological scenarios where causally-disconnected regions of the universe coexist and self-reproduce in some version of chaotic (eternal) inflation \cite{Linde1,Vilenkin1} (or a combination of both), has presented theoretical physicists with a new challenge. Traditionally, the goal of theoretical physics has been to develop explanations of natural phenomena based on predictive models. These models, in turn, are built using fundamental constants, which are measured in the lab to high accuracy. Such constants reflect the universality of the empirically-validated laws of physics as they apply to our observable universe, which I henceforth will characterize with a capital ``U''. Given this, many have expected that, as theories progressed, we would zero in on an over-arching explanation--the final theory--that would explain the values of these fundamental constants from first principles. This would be the crowning achievement of the reductionist program in physics, with roots that extend back all the way to the pre-socratic philosophers of ancient 
Greece\footnote{It would be naive to expect that this much-coveted final theory would have anything to say about complex systems, such as Earth's weather, living organisms, or the existence of consciousness. The reductionist program of high energy physics is limited to the fundamental components of matter and their interactions.}.

The aforementioned challenge to theoretical physics becomes relevant if the Universe is no longer unique. If there are many universes, either infinitely many or denumerably infinitely many, or even just more than one,  we must consider the possibility that each of these universes has a different set of fundamental constants and even of fundamental laws \cite{Weinberg1,Wilczek}. Here, physics enters unknown territory: we need not only theories that describe natural phenomena in terms of different sets of (untestable!) fundamental laws, but a theory that encompasses these theories--a theory of theories (TOT)--that provides a mechanism to discern between distinct choices of ``vacua'' wherein different kinds of physics take place.

Needless to say, we remain clueless as to how to build this TOT. There are essentially two choices: either we assume that the same set of fundamental laws apply across the multiverse and that only the values of the constants of nature can vary (i.e., each universe would have its own ``nature'' but they all share the same laws of physics), or we assume that there are no unique fundamental laws so that each universe can have its own set of fundamental laws and constants. The latter case is harder (currently impossible) to conceptualize. The former calls for selection effects, that is, for criteria based on certain assumptions that can reasonably restrict the choices of the fundamental parameters. For example, selection effects have been invoked to make sense of the small but finite value of the cosmological constant--which blatantly contradicts theoretical expectations from high energy physics \cite{Weinberg2, Vilenkin, Martel, GarrigaVilenkin}. Based on the existence of ``typical observers,'' with us as the only example, these selection effects illustrate current applications of weak anthropic reasoning in cosmology. Weinberg's prescient estimate of an upper bound $\rho_V^{\rm (max)}$ for $\rho_V$, the vacuum energy, used that in order for observers to exist galaxies and stars must form first \cite{Weinberg2}. So, it must be the case that $\rho_V^{\rm (max)} < \rho_{\rm matter}^{\rm (g.f.)}$, where  $\rho_{\rm matter}^{\rm (g.f.)}$ is the (baryon) matter density at the epoch of galaxy formation. If we use $t\sim 10^9$y for the time of galaxy formation, we obtain $\rho_V^{\rm (max)}\lesssim 1.37 \times 10^2 \rho_{\rm matter}^{\rm (0)}$, where $\rho_{\rm matter}^{\rm (0)}=0.046\rho_{\rm crit},$ according to the seven-year data from WMAP \cite{WMAP}. Thus, $\rho_V^{\rm (max)} < 8.63\rho_{\rm crit}$, in agreement with the present observational value of $\rho_V/\rho_{\rm crit} \simeq 0.7$. Had I used the density of baryonic plus dark matter, the bound worsens to $\rho_V^{\rm (max)} < 38.22\rho_{\rm crit}$.
This argument can be greatly refined, and the result narrowed to an upper bound much closer to the current value \cite{Martel,GarrigaVilenkin}.

The goal of the present work is not to delve into the merits and difficulties of multiverse scenarios or on the usefulness of anthropic reasoning in physics. There have been several works that address these issues, and some are listed in the references \cite{Ellis, Weinstein, MaorKraussStarkman,EllisSmolin}. Here, I wish to contrast statements made in cosmology, in particular those invoking ``typical observers'' or the ``principle of mediocrity,'' with current knowledge of the conditions to find life--and especially complex life--in planetary platforms. There is an enormous chasm, one could even say a two-culture disconnect, that must be avoided if anthropic reasoning is to really represent ``anthropic'' reasoning. As currently formulated, ``prebiotic principle'' would be more adequate.

\section{Typical Observers and The Principle of Mediocrity}

As formulated by Carter \cite{Carter,BarrowTipler}, the anthropic principle is more than a simple statement of the obvious: it proposes a different way of doing science, where inferences about the natural world can be made based on the existence of observers. The problem with most anthropic arguments is that what is commonly called ``observers'' should really be referred to as ``life.'' The conditions that must be satisfied for complex life to emerge in a planetary platform must be imposed {\it in addition} to the conditions for primitive, single-celled life. In what follows, I will assume that life is what we know it is, based on carbon biochemistry and capable of metabolism and reproduction in aqueous media. Certainly, there could be alternatives--life as we don't know it--with a silicon-based biochemistry and using ammonia as solvent, or even with a complete different and unanticipated genetics and biochemistry. If there are viable alternatives for life, we are far from understanding how \cite{MaorKraussStarkman}. So, we will restrict life to be a complex, self-sustaining chemical network capable of undergoing Darwinian evolution \cite{Joyce}. However, I note that there is no broadly accepted definition of life, and that some authors consider the task fundamentally misguided \cite{ClelandChyba}. 

Anthropic reasoning uses the fact that we exist--and by ``we'' it is meant intelligent observers--to place constraints on coupling constants and other cosmological parameters of physics. (For the purpose of this work, intelligent observers are beings capable of calling themselves intelligent.) As the examples in the next paragraph illustrate, the typical approach is to have a probability measure that depends on one or on a combination of fundamental couplings and cosmological parameters. This sort of reasoning may have something to say about the {\it consistency} requirements for primitive life (as we know it) to exist in a typical universe, {\it but has very little to say about the existence of complex life and nothing whatsoever to say about intelligent life}. For example, the range of the cosmological constant may be restricted, as done by Weinberg, Vilenkin, and collaborators \cite{Weinberg2,Vilenkin,Martel,GarrigaVilenkin}, or the inflaton potential may be extremely flat, as discussed by Vilenkin \cite{Vilenkin}, or the equation of state for dark energy may be restricted \cite{GarrigaLindeVilenkin}, or the matter-antimatter asymmetry and the range of density perturbations may be bounded, as done by Tegmark and 
Rees \cite{TegmarkRees}, or the Yukawa couplings of the Standard Model restricted, as done by Hogan \cite{Hogan}. (Note criticism by Weinberg \cite{Weinberg1}.) These selection criteria set the groundwork for the possible existence of primitive life in a given region of a universe, but have little to say about the emergence of complex life.

Here are a few illustrative examples. In 1995, Vilenkin proposed the ``principle of mediocrity'' (PM), whereby ``we are one of a infinite number of civilizations living in thermalized regions of the multiverse'' \cite{Vilenkin}. In this scenario, different inflating universes with fundamental constants within the right interval for civilizations to emerge are nucleated from a multiverse with probability 
\begin{equation}\label{prob}
d{\cal P}(\alpha)=Z^{-1}{\cal N}(\alpha)w_{\rm nucl}(\alpha)\prod_i d\alpha_i, 
\end{equation}
where $Z$ is a normalization factor, ${\cal N}(\alpha)$ is the average number of civilizations in such a universe (in its entire history), $w_{\rm nucl}(\alpha)\prod_i d\alpha_i$ is the nucleation probability for an inflating universe with $\alpha_i$ in the range $d\alpha_i$. In a later work, the PM was refined to ``our civilization is typical in the ensemble of all civilizations in the universe'' \cite{GarrigaVilenkin}, and the nucleation probability of eq. \ref{prob} was then adapted to reflect this locality, that is, our typicality in a universe abounding with civilizations. The probabilistic weight was chosen to be a certain range for the dark energy density--taken to be a random variable--so that $w_{\rm nucl}(\alpha)\prod_i d\alpha_i$ of eq. \ref{prob} became ${\cal P}_* (\rho_{\rm de}) d \rho_{\rm de}$: for civilizations to exist in different regions of the universe, the dark energy must have values within a certain range, that is, it must allow for galaxies and stars to form and to exist long enough, limiting redshifts to $z\lesssim 1$. 

More recently, Bousso et al. have argued, in the context of eternally inflating bubbles, that observers that live at a time $t_{\rm obs}$ after the formation of their vacuum bubble are likely to measure a cosmological constant of order 
$\Lambda \sim t^{-2}_{\rm obs}$ \cite{Boussoetal,BoussoHarnik}. The authors then state that this condition is quite general, independent of the nature of the ``observers,'' that is, that they do not need to be ``like us.'' A general principle was then proposed, the ``entropic principle,'' whereby entropy production is used as a proxy for the existence of complex organic structures: in a given environment, no ``observers'' are produced without an increase in entropy \cite{BoussoHarnik}. There follows an inference that the number of observers is proportional, on average, to the net entropy produced, $n_{\rm obs} \propto \Delta S$. The authors are careful to state that entropy production or the existence of galaxies is not sufficient for the existence of observers. According to them, and this is a key point, the advantage of such general principle is that it can be used within the limited framework of field theory and cosmology, where organic molecules and planetary systems cannot be properly accounted for. Herein lies the problem. No cosmological argument using the existence of large mass galaxies at low redshift and/or the production of entropy in stellar systems is a guarantee of complex, multicellular life; at most it could serve as a necessary precondition for primitive life, and only in planetary platforms conducive to life. The word ``observer'' is used out of context, adding much confusion to the field and creating very negative philosophical repercussions. The cosmological and astrophysical selection criteria quoted in the literature have little to say about the existence of intelligent life forms. They can't. There is an urgent need to disentangle the conditions for the emergence of life from those for the emergence of complex and intelligent life.

\section{Conditions for the Emergence of Complex Life} 

In the early 1960s, Drake proposed an equation to estimate the probability of having technologically-advanced civilizations in our galaxy. His strategy was to break the problem down into its many factors following a simple hierarchy: stars with planets, planets with life, planets with intelligent life, and so on \cite{Drake}. In the spirit of the Drake equation, we first establish a simple hierarchical subdivision to help disentangle the different sets of conditions that must be imposed on the (long and winding) road from the multiverse to complex life. Start with an ensemble of universes which may be the vacua of a string landscape multiverse or causally-disconnected nucleating bubbles in an eternally-inflating cosmos, ${\cal U}  = \{ u_i\}$, where $i=1,\dots N \leq \infty$ labels different realizations according to the set $\{c_j\}$ of fundamental constants and cosmological parameters in each\footnote{There is a serious conceptual barrier here, since we don't have a probabilistic measure in the multiverse. Not knowing how to compute the probability amplitude of a nucleating universe with certain properties, all we can do is speculate \cite{MaorKraussStarkman}. Note also that if the number of universes is infinite, there should exist infinitely many copies of our own Universe. Ellis and Smolin recently argued that within the landspace, given that the presently known number of vacua with $\Lambda < 0$ is much larger than that with $\Lambda > 0$, the probability of a universe with $\Lambda > 0$ is negligible \cite{EllisSmolin}.}. Consider the subset of ${\cal U}$ that contains the universes where these constants and parameters satisfy anthropic (or better, prebiotic) bounds, $\{ u_{\rm cosmo}\}
\subset \{ u_i\} $. Our observable Universe ($U$) belongs, of course, to this set ($U \in \{ u_{\rm cosmo}\}$). Let us call these necessary but not sufficient conditions the ``cosmological constraints on life,'' and denote them by $\{c_{\rm cosmo}\} \subset\{ c_j\}$. To this, we must add three other sets of conditions: i. {\it astrophysical conditions} ($\{c_{\rm astro}\}$) will ensure that large enough galaxies and the right kinds of stars (more later) will form within a subset of universes belonging to $\{ u_{\rm cosmo}\}$; ii. {\it planetary conditions} ($\{c_{\rm life}\}$) will ensure that planets (and moons) with the right properties to harbor life will form around the right kinds of stars; iii. {\it complex life conditions} ($\{c_{\rm complex~life}\}$) will ensure that a number of these planets will have the right conditions for complex life to emerge. Taking all of these together, the conditions for complex life to emerge in a particular universe out of a multiverse must satisfy,
\begin{equation}\label{conditions}
\{c_{\rm cosmo}\} \oplus \{c_{\rm astro}\} \oplus \{c_{\rm life}\} \oplus \{c_{\rm complex~life}\},
\end{equation}
where the symbol $\oplus$ here denotes ``together with''. There is an implicit hierarchical ordering from left to right: $\{c_{\rm complex~life}\}$ can only be satisfied if $\{c_{\rm life}\}$ is satisfied and so on, but {\it not} vice-versa. Each set of conditions deals with different science: from still vague selection procedures in the multiverse to high metallicity stellar environments to planetary environments which are fairly stable for billions of years. A Drake-like equation connecting the multiverse to the existence of complex life in a universe would attribute a probability to each of the four sets above, so that the fraction of nucleating universes that can harbor complex life is equal to the fraction of universes that satisfy the cosmological constraints multiplied by the fraction that also satisfies the astrophysical, planetary, and complex life constraints,
\begin{equation}\label{Drake1}
{\cal F}_{\rm ucl} = f_{\rm cosmo}f_{\rm astro}f_{\rm life}f_{\rm complex~life}.
\end{equation}
Of course, a quantitative formulation of such equation would need probabilistic measures for each of the sets of conditions which, as in the case of the original Drake equation \cite{Drake}, are either difficult to estimate with precision or are lacking.

A last step would be to attribute a set of conditions for intelligent life to emerge out of a subset of planets bearing complex life forms. However, as the history of life on Earth teaches us, we have no reason to expect that intelligent life is a necessary consequence of complex, multi-cellular life. Evolution does not imply in multicellularity or in intelligence, only in well-adapted life forms. As 150 million years of low-functioning dinosaurs demonstrate, we should not connect the existence of complex life with that of intelligent observers. Thus, a fifth and last entry on our list of conditions remains unknown, as of course was also the case with the original Drake equation \cite{Drake} and its more recent formulations \cite{Drake2}.

To establish even a preliminary list of relevant physical parameters for each conditional set is a major challenge. To establish their range is another. However, we can make a few statements about each of them:\\
\noindent$\bullet~$ The first set, $\{c_{\rm cosmo}\}$, has been the object of much activity in the cosmological community. A few obvious parameters, as we have seen above, are: vacuum energy density, matter-antimatter asymmetry, dark matter density, and perturbations for the formation of large-scale structure. To these, one adds the couplings of the four interactions and the masses of quarks and leptons so that hadrons and then nuclei can form after electroweak symmetry breaking. I'd also add universes with small initial entropy $S_i$ \cite{Carroll}, since large-entropy initial states will not be conducive of structure formation of any kind: for complexity to emerge there must exist enough free energy \cite{Entropy}.\\
\noindent$\bullet~$ The second set, $\{c_{\rm astro}\}$, includes constraints in galactic morphology and stellar ages and types. In order to retain heavy elements, galaxies must have masses above a certain value. Numerical results suggest that galaxies with $M\geq 10^9M_{\odot}$ are able to retain about $\geq 30\%$ of their heavy elements \cite{LowFerrara}. Merging processes to form galactic disks such as that of the Milky Way (which has a mass $M_{\rm MW}\sim 10^{12}M_{\odot}$), take time to complete and are unlikely for $z > 1$ \cite{AbrahamBergh}. So, life is favored in large-mass galaxies, and we could take $M_{\rm MW}$ as a fiducial value. Stars can be constrained by type, since continually habitable zones (i.e., surface liquid water for extended periods) exist only around stars in the spectral classes between F5 to mid-K \cite{Kasting1,Livio}, which amount to about $\sim 20\%$ of main sequence stars. Of these, we must include only the fraction bearing planetary systems. As indicated by current searches for exoplanets, this fraction can be substantial. However, we should discount stars in binary or multiple systems, about $2/3$ of the total number of stars, which may not have planets at all. I will leave the notion of a galactic habitable zone aside.\\
\noindent$\bullet~$ The third set, $\{c_{\rm life}\}$, includes the planetary and chemical constraints for the existence of life. Of the planets in the habitable zone, only a fraction will have the right preconditions for life: liquid water and the elements C, O, H, N, and the less abundant but no less needed P, S, Fe, Ca, Na, Cl, etc. Apart from water, other simple molecules are also supposed to be present, although the specifics may differ (CH$_4$, CO$_2$, NH$_3$...). As indicated by Miller-type experiments \cite{Miller}, to produce amino acids {\it in situ} there is also a need for a reducing atmosphere. Planetary platforms with volcanism have a clear advantage \cite{Bada}. (It is possible that amino acids and other organics drifted to Earth or were brought here via meteoritic impacts \cite{Pizzarello}, helping to jump-start the prebiotic chemistry that led to abiogenesis.) If the history of life on Earth serves as a model, we know of three important facts: first, that life can emerge fairly quickly after a planet reaches a certain level of crustal quiescence. If the last great bombardment ended 3.9 Gya (billion years ago) \cite{Bombardment} and the first signs of life are dated at 3.5 Gya \cite{Buick}, the window is of only a few hundred million years \cite{Orgel,Gleiser}. Davies and Lineweaver have claimed that the window could be even narrower and that life may have emerged several times until it finally took hold sometime around 3.5 Gya \cite{Davies}. In this case, simple life could presumably be widespread in the cosmos. The second fact is that life remained single-celled for about 2.5 billion years: the first multicellular organisms that we know of (sponges are hard to fossilize) date from about 1 Gya or less. If life is widespread, it will be mostly composed of single-celled organisms. Third, a plethora of extremophiles discovered over the past decades illustrates how versatile and resilient single-celled life can be \cite{Barossetal}. So, if single-celled life takes hold, natural selection will lead to adaptability in the most diverse environments.\\
\noindent$\bullet~$ The fourth set, $\{c_{\rm complex~life}\}$, includes planetary constraints that would essentially ensure the continuity of fairly stable environmental conditions thought to be needed for complex, multicellular life to evolve. To begin, terrestrial life's complexity seems to be intrinsically linked to the rise of atmospheric oxygen, courtesy of photosynthetic blue-green algae. To date, there is only one known fully anaerobic multicellular organism (a microscopic nematode found deep in the Black Sea \cite{NickLane}); some worms may switch between aerobic and anaerobic respiration \cite{Schottler}, while deep-vent tube worms are still aerobic, apparently symbiotic with anaerobic bacteria that supply them with sulfide \cite{Martin}. Anoxic environments don't seem very conducive to multicellular life, although there is much we don't know of Earth's biota. However, we do know that complex life is more fragile and requires more efficient energy-processing metabolism to be sustained: planetary platforms able to harbor complex life-forms must thus satisfy extra requirements. Some that have been suggested include \cite{WardBrownlee}: i. a magnetic field strong enough to protect the surface from damaging cosmic radiation and, for small planets, to keep the atmosphere from being sputtered away (as happened to Mars); ii. plate tectonics that work as a global thermostat, recycling chemicals that help regulate the levels of carbon dioxide and keep the global temperature stable. Water can then remain liquid for billions of years. If not tectonics, other global temperature-stabilizing mechanism is needed; iii. a large moon to stabilize the planet's tilt. Without it, there would be no regular seasons and habitability would be severely compromised \cite{Laskar}. Other factors have been proposed (e.g. the presence of a large planet on the outskirts of the stellar system to protect inner rocky planets from excessive asteroid and cometary impacts, as Jupiter does for us), but these will suffice.

\section{Discussion}

It is often said that proponents of the anthropic principle have some crypto-teleological agenda, that the principle implies some kind of supernatural agency in the cosmos. Although this may be the case for the strong version of the principle (the universe in a sense created us), the opposite is true of the weak version: in the context of a multiverse, where every possible universe may exist, ours is not special; it just so happens to have the properties that allow for us to be here \cite{Weinberg1}. There are arguments going back and forth among colleagues as to whether the ``predictability'' of anthropic arguments is of any use. The point of the present work is not to engage in such arguments. (For that, see Ref. \cite{Tear}.) My main goal here is to address what I believe is a widespread misconception, that the conditions usually imposed in anthropic arguments have something to say about our existence or typicality in the Universe. Accordingly, I have argued that general cosmological conditions on the values of the vacuum energy, the masses and couplings of the fundamental particles and interactions, entropic arguments, or the presence of large-mass galaxies, have nothing to say about the existence of complex, multicellular life. At best, such selection criteria set the preconditions for life and should be called ``prebiotic'' instead of ``anthropic.'' In fact, it is perfectly plausible that a universe satisfying all of these conditions will be devoid of any form of complex life and even more so of intelligent observers. Insisting that such conditions guarantee our presence in this Hubble volume overlooks the evolutionary complexities of multicellular life forms. It implicitly--and erroneously--assumes that given enough time primitive life evolves into intelligent life. As presently construed, cosmological and physical arguments can only go so far along the road toward life; the existence of complex, multicellular life is not covered.

If what we have learned thus far about our solar system and other planetary systems is any indication, Earth-like planets are not the rule: continuously habitable zones are narrow and restricted to a minority of stars. That said, NASA's Kepler mission will hopefully zero in on a few worthy candidates and, if present expectations are vindicated, we should even be able to have rudimentary knowledge of their atmospheric composition \cite{Kaltenegger}. Ground-based techniques are also producing results \cite{Rogers}. Remaining optimistic, it is indeed possible that primitive life forms, perhaps not unlike our own archaea and bacteria, will be present in extraterrestrial environments, benign or even extreme. However, even if we should always be open to surprises, the existence of complex organisms--and even more so of intelligent organisms--is, in all probability, exceedingly rare, a point Carter also made \cite{Carter}. We may not be the only intelligent species in the Universe, or even in the Milky Way. But it is very hard to argue at present that we are a typical sampling of a vast roster of extraterrestrial civilizations.

\section*{Acknowledgments}

This work is supported in part by a National Science Foundation Grant No. PHYS-0757124. I thank Ursula Goodenough, Nick Lane and William Martin for information on anaerobic complex life, and Cesar Zen Vasconcellos for the invitation to contribute to the proceedings of the 2009 International Workshop on Astronomy and Relativistic Astrophysics.


\section{References}

\begin{thebibliography}{00}

\bibitem{Susskind} R. Bousso and J. Polchinski, {\it JHEP} {\bf 6} (2000) 6;
S. Kachru et al. {\it Phys. Rev.} D{\bf 68} (2003) 046005; L. Susskind, {\it The Anthropic Landscape of String Theory}, arXiv:hep-th/0302219 (2003); {\it ibid.}, {\it The Cosmic Landscape: String Theory and the Illusion of Intelligent Design} (Little Brown, New York, 2006).

\bibitem{Linde1} A. D. Linde, {\it Phys. Lett.} B{\bf 129} (1983) 177. A general overview of Linde's many contributions can be found in A. Linde, Inflation, Quantum Cosmology, and the Anthropic Principle, in {\it Science and Ultimate Reality}, eds. J. D. Barrow, P. C. W. Davies, and C. L. Harper, Jr. (Cambridge University Press, Cambridge, 2004).

\bibitem{Vilenkin1} A. Vilenkin, {\it Phys. Rev.} D{\bf 27} (1983) 2848; A. D. Linde, {\it Phys. Lett.} B{\bf 175} (1986) 395.

\bibitem{Weinberg1} S. Weinberg, Living in the Multiverse, in {\it Universe or Multiverse?}, ed. B. Carr (Cambridge University Press, Cambridge, 2007).

\bibitem{Wilczek} F. Wilczek, Enlightenment, Knowledge, Ignorance, Temptation, in {\it Universe or Multiverse?}, ed. B. Carr (Cambridge University Press, Cambridge, 2007).

\bibitem{Weinberg2} S. Weinberg, {\it Phys. Rev. Lett.} {\bf 59} (1987) 2607.

\bibitem{Vilenkin} A. Vilenkin, {\it Phys. Rev. Lett.} {\bf 74} (1995) 846.

\bibitem{Martel} H. Martel, P. R. Shapiro, and S. Weinberg, {\it Astrophys. J.} {\bf 492} (1998) 29.

\bibitem{GarrigaVilenkin} J. Garriga and A. Vilenkin, {\it Phys. Rev.} D{\bf 61} (2000) 083502; {\it ibid.}, D{\bf 64} (2001) 023517; {\it ibid.}, D{\bf 67} (2003) 043503.

\bibitem{WMAP} N. Jarosik et al., arXiv:astro-ph/1001.4744 (2010).

\bibitem{Ellis} G. F. R. Ellis, U. Kirchner, and W. Stoeger, {\it Mon. Not. Roy. Ast. Soc.} {\bf 347} (2004) 921; L. Smolin, {\it Life of the Cosmos} (Oxford University Press, New York, 1997).

\bibitem{Weinstein} S. Weinstein, {\it Class. Quant. Grav.} {\bf 23} (2006) 4231.

\bibitem{MaorKraussStarkman} I. Maor, L. Krauss, and G. Starkman, {\it Phys. Rev. Lett.}
{\bf 100} (2008) 041301.

\bibitem{EllisSmolin} G. F. R. Ellis and L. Smolin, arXiv:hep-th/0901.2414.

\bibitem{Carter} B. Carter, Confrontation of Cosmological Theories with Observational Data, in {\it Proceedings of the I. A. U. Symposium {\bf 63}}, ed. M. Longair (Reidel, Dordrecht, 1974); {\it ibid.}, {\it Philos. Trans. R. Soc. London} A{\bf 310} (1983) 347.

\bibitem{BarrowTipler} J. D. Barrow and F. J. Tipler, {\it The Anthropic Cosmological Principle} (Oxford University Press, Oxford, 1986).

\bibitem{Joyce} G. F. Joyce, Forward, in {\it Origins of Life: The Central Concepts}, eds. D. W. Deamer and G. R. Fleischaker (Jones \& Bartlett, Boston, 1994).

\bibitem{ClelandChyba} C. E. Cleland and C. F. Chyba, {\it Orig. Life Evol. Biosp.} {\bf 32} (2002) 387.

\bibitem{GarrigaLindeVilenkin} J. Garriga, A. D. Linde, and A. Vilenkin, {\it Phys. Rev.} D{\bf 69} (2004) 063521.

\bibitem{TegmarkRees} M. Tegmark and M. J. Rees, {\it Astrophys. J.} {\bf 499} (1998) 526.

\bibitem{Hogan} C. Hogan, {\it Rev. Mod. Phys.} {\bf 72} (2000) 1149.

\bibitem{Boussoetal} R. Bousso et al., {\it Phys. Rev.} D{\bf 76} (2007) 043513.

\bibitem{BoussoHarnik} R. Bousso and R. Harnik, {\it The Entropic Landscape}, arXiv:hepth/1001.1155.

\bibitem{Drake} F. D. Drake, {\it Intelligent Life in Space} (Macmillan, New York, 1962).

\bibitem{Drake2} C. Walters, R. A. Hoover, and R. K. Kotra, {\it Icarus} {\bf 41} (1980) 193.

\bibitem{Carroll} S. M. Carroll and J. Chen, {\it Int. J. Mod. Phys.} D{\bf14} (2005) 2335.

\bibitem{Entropy} I. Prigogine, {\it Introduction to Thermodynamics of Irreversible Processes} (John Wiley \& Sons, New York, 1967).

\bibitem{LowFerrara} M. Mac Low and A. Ferrara, {\it Astrophys. J.} {\bf 513} (1999) 142.

\bibitem{AbrahamBergh} R. G. Abraham and S. van der Bergh, {\it Science} {\bf 293} (2001) 1273.

\bibitem{Kasting1} J. F. Kasting, D. P. Whitmire, and R. T. Reynolds, {\it Icarus} {\bf 101} (1993) 108.

\bibitem{Livio} M. Livio, {\it Astroph. J.} {\bf 511} (1999) 429.

\bibitem{Miller} S. L. Miller, {\it Science} {\bf 117} (1953) 528; A. Lazcano and J. L. Bada, {\it Orig. Life Evol. Biosp.} {\bf 33} (2004) 235.

\bibitem{Bada} A. P. Johson et al., {\it Science} {\bf 322} (2008) 404.

\bibitem{Pizzarello} J.~R. Cronin and S. Pizzarello,  {\it Geochim. Cosmochim. Acta} {\bf 50} (1986) 2419; {\it ibid.} {\it Science} {\bf 275} (1997) 951.

\bibitem{Bombardment} See, e.g., R. Gomes et al., {\it Nature} {\bf 435} (2005) 466.

\bibitem{Buick} R. Buick, The Earliest Records of Life on Earth, in {\it Planets and Life}, eds. W. T. Sullivan III and J. A. Baross (Cambridge University Press, Cambridge, 2007).

\bibitem{Orgel} L.~E. Orgel, {\it Orig. Life Evol. Biosph.} {\bf 28} (1998) 91.

\bibitem{Gleiser} An illustration of how this window can impact studies of the origin if life can be found in M. Gleiser, {\it Orig. Life. Evol. Biosph.} {\bf 37} (2007) 235.

\bibitem{Davies} P. C. W. Davies and C. H. Lineweaver, {\it Astrobiology} {\bf 5} (2005) 154.

\bibitem{Barossetal} J. A. Baross, J. A. Huber, and M. O. Schrenk, Limits of Carbon Life on Earth and Elsewhere,  in {\it Planets and Life}, eds. W. T. Sullivan III and J. A. Baross (Cambridge University Press, Cambridge, 2007).

\bibitem{NickLane} N. Lane, private communication.

\bibitem{Schottler} U. Sch\"ottler, {\it J. Comp. Physiol.} {\bf 125} (1978) 185.

\bibitem{Martin} See, e.g., A. G. Tielens et al. {\it Trends Biochem. Sci.} {\bf 27} (2002) 564, for anaerobic mitochondria.

\bibitem{WardBrownlee} For a summary see, P. D. Ward and D. Brownlee, {\it Rare Earth: Why Complex Life Is Uncommon in the Universe} (Springer, New York, 2000). 

\bibitem{Laskar} J. Laskar, F. Joutel, and P. Robutel, {\it Nature} {\bf 361} (1993) 615.

\bibitem{Tear} M. Gleiser, {\it A Tear at the Edge of Creation: A Radical New Vision for Life in an Imperfect Universe} (Free Press, New York, 2010); {\it ibid.} {\it The Prophet and the Astronomer: A Scientific Journey to the End of Time} (W. W. Norton, New York, 2002).

\bibitem{Kaltenegger} L. Kaltenegger and J. Kasting, {\it Astrobiology} {\bf 8} (2008) 394.

\bibitem{Rogers} J. C. Rogers et al., arXiv:0910.1257 (in press, {\it Astrophys. J.}).

\end{thebibliography}
\end{document}